\documentclass[9pt,technote]{IEEEtran} %

\usepackage{cite}
\ifCLASSINFOpdf
\else
\fi
\usepackage{amsmath}
\usepackage{url}
\usepackage{graphicx}
\usepackage{amsmath, amssymb}
\usepackage{bm}

\usepackage[font={small,it}]{caption}
\usepackage{color}

\usepackage{hyperref} %

\usepackage{algorithm}
\usepackage[noend]{algpseudocode}

\usepackage{hhline}
\usepackage{array}
\usepackage{stackengine}

\usepackage{xspace}

\newcommand\mydots{\hbox to 1em{.\hss.\hss.}}

\newcommand{\dd}{\textrm{d}}
\newcommand{\dt}{\textrm{dt}}

\newcommand{\udim}{m}

\newcommand{\N}{\mathcal{N}}

\newcommand{\X}{\mathcal{X}}

\newcommand{\R}{\mathbb{R}}
\newcommand{\Prob}{\mathbb{P}}

\begin{document}
\title{On the Problem of Reformulating Systems with Uncertain Dynamics as a Stochastic Differential Equation}

\vspace{-8mm}

\author{Thomas Lew, Apoorva Sharma, James Harrison,\\Edward Schmerling, Marco Pavone
\thanks{The authors are with the Department of Aeronautics \& Astronautics, Stanford University, Stanford, CA 94305-4035 USA (emails: \texttt{\{thomas.lew, apoorva, jharrison, schmrlng, pavone\}@stanford.edu}).}
}

\maketitle
\begin{abstract}
We identify an issue in recent approaches to  learning-based control that reformulate systems with uncertain dynamics using a stochastic differential equation. 
Specifically, we discuss the approximation that replaces a model with fixed but uncertain parameters (a source of epistemic uncertainty) with a model subject to external disturbances modeled as a Brownian motion (corresponding to aleatoric uncertainty).
\end{abstract}

\vspace{-4mm}

\IEEEpeerreviewmaketitle

\section{Problem Formulation and Error in Literature}
Consider a nonlinear system whose state at time $t\geq 0$ is $x(t)\in\R^n$, control inputs are $u(t)\in\R^\udim$, such that
\begin{align}\label{eq:dynamics}
    \dot x(t) = f(x(t),u(t)),
    \quad t\in[0,T],
\end{align}
where $T>0$, $x(0)=x_0\in\R^n$ almost surely, i.e., $x(0)$ is known exactly, and $f:\R^n\times\R^m\rightarrow\R^n$ is twice continuously differentiable.

In many applications, $f$ is not known exactly, and prior knowledge is necessary to safely control \eqref{eq:dynamics}. One such approach consists of assuming that $f$ lies in a known space of functions $\mathcal H$, and to impose a prior distribution in this space $\Prob(\mathcal H)$. For instance, by assuming that $f$ lies in a bounded reproducing kernel Hilbert space (RKHS), a common approach consists of imposing a Gaussian process prior on the uncertain dynamics $f\sim\mathcal{GP}(m,k)$, where $m:\R^{n+m}\,{\rightarrow}\,\R^n$ is the mean function, and $k:\R^{n+m}\times\R^{n+m}\,{\rightarrow}\,\R^{n\times n}$ is a symmetric positive definite covariance kernel function which uniquely defines $\mathcal{H}$ \cite{williams2006gaussian,Alvarez2012}. 
An alternative consists of assuming that $f(x,u)=\phi(x,u)\theta$, where $\phi:\R^{n+m}\rightarrow\R^{n\times p}$ are known basis functions, and $\theta\in\R^p$ are unknown parameters. With this approach, one typically sets a prior distribution on $\theta$, e.g., a Gaussian $\theta\sim\N(\bar\theta,\Sigma_\theta)$, and updates this belief as additional data about the system is gathered.

Given these model assumptions and prior knowledge about $f$, safe learning-based control algorithms often consist of designing a control law $u$ satisfying different specifications, e.g., minimizing fuel consumption $\|u\|$, or satisfying constraints $x(t)\in\mathcal{X}\ \forall t\in[0,T]$, with $\mathcal{X}$ a set encoding safety and physical constraints.

Next, we describe an issue with the mathematical formulation of the safe learning-based control problem that has appeared in recent research  \cite{Chowdhary2015,dfan2020ICRA,Nakka2020,Joshi2021}, slightly changing notations and assuming a finite-dimensional combination of features for clarity of exposition but without loss of generality. As in \cite{Nakka2020}, consider the problem of safely controlling the uncertain system
\begin{align}\label{eq:dynamics:original}
    \dot x(t) = \phi(x(t),u(t))\theta,
    \quad
    \theta\sim\N(\bar\theta,\Sigma_\theta),
\end{align}
where 
    $t\in[0,T]$,  $x(0)=x_0$,  $\bar\theta\in\R^p$, and $\Sigma_\theta\in\R^{p\times p}$ is positive definite, with $\Sigma_\theta=B_\theta B_\theta^{\top}$ its Cholesky decomposition. Note that this formulation can be equivalently expressed in function space, where $f$ is drawn from a Gaussian process with mean function $m(x,u) = \phi(x,u)\bar\theta$ and kernel $k(x,u,x',u') = \phi(x,u) \Sigma^{-1}_\theta \phi(x',u')^{\top}$. Our representation can be seen as a weight-space treatment of the GP approaches used in \cite{Chowdhary2015} and \cite{dfan2020ICRA}.\footnote{For a squared exponential kernel, one needs $p\rightarrow\infty$ for this equivalence, see \cite{williams2006gaussian} for more details. Nevertheless, the issue discussed in this paper remains valid for such kernels.}

These works then proceed by introducing the Brownian motion $W(t)$, making the change of variable $\theta\dd t=\bar\theta\dd t + B_\theta\dd W(t)$, and reformulating \eqref{eq:dynamics:original} as a stochastic differential equation (SDE)
\begin{align}\label{eq:dynamics:reformulated}
    \dd x(t) = \phi(x(t),u(t))\bar\theta\dd t + 
    \phi(x(t),u(t)) B_\theta\dd W(t),
\end{align}
with $t\in[0,T]$ and $x(0)=x_0$. 
Unfortunately, \eqref{eq:dynamics:reformulated} is not equivalent to \eqref{eq:dynamics:original}. Indeed, the solution to \eqref{eq:dynamics:reformulated} is a Markov process, whereas the solution to \eqref{eq:dynamics:original} is not. %
Intuitively, the increments of the Brownian motion $W(t)$ in \eqref{eq:dynamics:reformulated} are independent, whereas in \eqref{eq:dynamics:original}, $\theta$ is randomized only once, and the uncertainty in its realization is propagated along the entire trajectory. 
By making this change of variables for $\theta\dd t$, the temporal correlation between the trajectory $x(t)$ and the uncertain parameters $\theta$ is neglected. 
In the next section, 
we provide a few examples to illustrate the distinction between these two cases. The first demonstrates the heart of the issue on a simple autonomous system, whereas the second shows that analyzing the SDE reformulation \eqref{eq:dynamics:reformulated} is insufficient to deduce the closed-loop stability of the system in \eqref{eq:dynamics:original}.

\section{Counter-Examples}
\subsection{Uncontrolled system}
Consider the scalar continuous-time linear system
\begin{align}
    \dot x(t) &= \theta, 
    \quad \theta\sim\N(0,1), 
    \quad t\in [0,T],
    \label{eq:ex:scalar_orig}
\end{align}
where $T>0$ and $x(0)=0$ almost surely, i.e., $x(0)$ is known exactly. 
The solution to \eqref{eq:ex:scalar_orig} satisfies $x(t)=\theta t$, i.e. each sample path is a linear (continuously differentiable) function of time $t\in[0,T]$. The marginal distribution of this stochastic process is Gaussian at any time $t\in[0,T]$, with  $x(t)\sim\N(0,t^2)$. 
The increments of this process are not independent, since the increment 
$x(t_2)-x(t_1)=(t_2/t_1)x(t_1)$ depends on $x(t_1)-x(0)=x(t_1)$ for any $t_2>t_1>0$. 

\begin{figure}
    \includegraphics[width=0.805\linewidth]{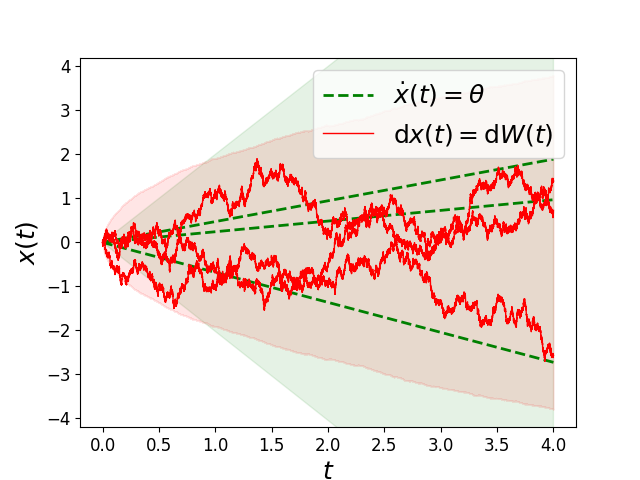}
    \caption{Visualization of sample paths and confidence intervals for the open loop system given by \eqref{eq:ex:scalar_orig} (green) and for the reformulation presented in \cite{Chowdhary2015,dfan2020ICRA,Nakka2020,Joshi2021} (red): the solutions 
    of \eqref{eq:ex:scalar_orig} and of \eqref{eq:ex:scalar_wrong} are distinct. 
    The solid and dashed lines represent a handful of sample paths. The shaded regions represent the marginal $95\%$ confidence intervals. }
    \label{fig:params_uncertainty_problem1}
\end{figure}

Using the change of variables described previously, 
one might consider substituting $\dd W(t)$ for $\theta\dt$, 
where $W(t)$ is a standard Brownian motion, yielding the following SDE
\begin{align}
    \dd x(t)  = \dd W(t), 
    \quad x(0)\,{=}\,0 \ \text{(a.s.)}
    \quad t\in [0,T].
    \label{eq:ex:scalar_wrong}
\end{align}
The solution of this SDE is a standard Brownian motion $x(t)\,{=}\,W(t)$ started at $W(0)\,{=}\,0$. 
This stochastic process has different marginal distributions $x(t)\sim\mathcal{N}(0,t)$, 
has independent increments, 
and is not differentiable at any $t$ almost surely. 
We illustrate sample paths of these two different stochastic processes in Figure \ref{fig:params_uncertainty_problem1}.

\subsection{System with linear feedback}

\begin{figure}
    \includegraphics[width=0.75\linewidth,trim=30 0 0 0, clip]{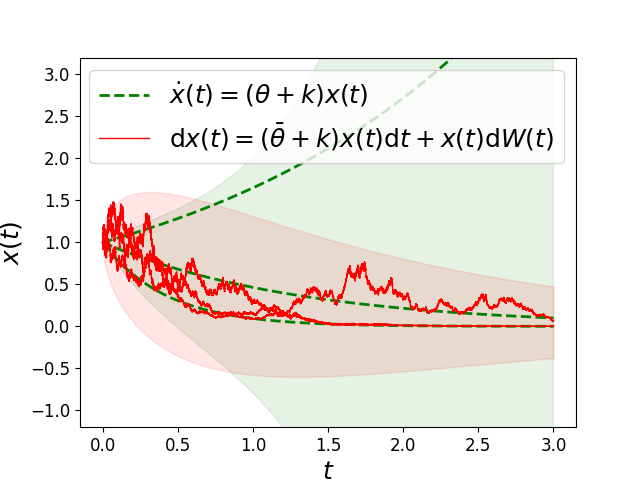}
\centering
    \caption{
    Visualization of sample paths and confidence intervals for the controlled system given by \eqref{eq:controlled:scalar_orig} (green) and for the reformulation presented in \cite{Chowdhary2015,dfan2020ICRA,Nakka2020,Joshi2021} (red): the solutions 
    of \eqref{eq:controlled:scalar_orig} and of \eqref{eq:controlled:scalar_wrong} are distinct. 
    The solid and dashed lines represent a handful of sample paths. The shaded regions represent the marginal $95\%$ confidence intervals. }
    \label{fig:params_uncertainty_problem2}
\end{figure}

Starting from $x(0)=x_0\in\R$, consider the controlled linear system
\begin{align}
    \dot x(t) &= \theta x(t) + u(t)
    =
    (\theta+k)x(t), 
    \ \theta\sim\N(\bar\theta,1), 
    \ t\in [0,T],
    \label{eq:controlled:scalar_orig}
\end{align}
where $k\in\R$ is a feedback gain and $u(t)=kx(t)$ is the state-feedback control policy. %
Solutions to \eqref{eq:controlled:scalar_orig} take the form $x(t)=x_0e^{(\theta+k)t}$. Choosing the gain $k=-(\bar\theta+1)$ and simulating from $x_0=1$, one obtains the sample paths shown in Figure \ref{fig:params_uncertainty_problem2}. We observe that some sampled trajectories are unstable, %
corresponding to samples of $\theta$ such that $\theta+k>0$.

Note that the substitution $\theta\dt =\bar\theta\dt + \dd W(t)$ %
yields the SDE
\begin{align}
    \dd x(t)  = (\bar\theta+k)x(t)\dd t + x(t)\dd W(t), 
    \quad t\in [0,T].
    \label{eq:controlled:scalar_wrong}
\end{align}
The solution of this SDE is a geometric Brownian motion $x(t)=x_0e^{((\bar\theta+k-\frac{1}{2})t+W(t))}$. Choosing the same control gain $k=-(\bar\theta+1)$ and plotting sample paths in Figure \ref{fig:params_uncertainty_problem2}, we observe that the system \eqref{eq:controlled:scalar_wrong} is stochastically stable.

\subsection{Discrete-time system}
The observation we make in this note is well-known in the discrete-time problem setting. For example, starting from $x_0\in\R$, the linear system with multiplicative uncertainty
\begin{align}\label{eq:example_dt}
    x_{t+1} &= \theta x_t, 
    \quad \theta\sim\N(\bar\theta,1), 
    \quad t\in \mathbb{N},
\end{align}
is different from the system with additive disturbances
\begin{align}\label{eq:example_dt_w}
    x_{t+1} &= \bar\theta x_t+w_k, 
    \quad w_k\sim\N(0,1), 
    \quad t\in \mathbb{N},
\end{align}
where the disturbances $(w_k)_{k\in\mathbb{N}}$ are independent and identically distributed.
We refer to \cite[Chapter 4.7]{McHutchon2014} for further discussions about this topic. We also refer to \cite{SmithCSL21} for a recent analysis of systems of the form of \eqref{eq:example_dt} where the parameters $\theta$ are resampled at each time $t$.  

\section{Implications and Possible Solutions} 
As \eqref{eq:dynamics:original} and \eqref{eq:dynamics:reformulated} are generally not equivalent, the stability and constraint satisfaction guarantees derived for the SDE \eqref{eq:dynamics:reformulated} in recent research  \cite{Chowdhary2015,dfan2020ICRA,Nakka2020,Joshi2021} do not necessarily hold for  the system \eqref{eq:dynamics:original}.\footnote{For instance, the generator of the Markov process solving the SDE \eqref{eq:dynamics:reformulated} (see \cite{LeGall2016}) is used to prove stability in \cite{Chowdhary2015,dfan2020ICRA,Nakka2020,Joshi2021}. Unfortunately, \eqref{eq:dynamics:original} does not yield a Markov process. Thus, it would be necessary to adapt the concept of generator to solutions of \eqref{eq:dynamics:original} before concluding the stability of the original system.} This could yield undesired behaviors when applying such algorithms, developed on an SDE formulation of dynamics \eqref{eq:dynamics:reformulated}, to safety-critical systems where uncertainty is better modeled by \eqref{eq:dynamics:original}, i.e., dynamical systems with uncertain parameters that are not changing over time.

Although \eqref{eq:dynamics:reformulated} is not equivalent to \eqref{eq:dynamics:original}, it is interesting to ask whether \eqref{eq:dynamics:reformulated} is a conservative reformulation of \eqref{eq:dynamics:original} for the purpose of \textit{safe} control. 
For instance, given a safe set $\X\subset\R^n$, if one opts to encode safety constraints through joint chance constraints of the form
\begin{align}\label{eq:cc}
    \Prob(x(t)\in\mathcal X \ \forall t\in[0,T])\geq (1-\delta),
\end{align} where $\delta\in(0,1)$ is a tolerable probability of failure, there may be settings where a controller $u$ satisfying 
\eqref{eq:cc} for the SDE \eqref{eq:dynamics:reformulated} 
may provably satisfy 
\eqref{eq:cc} for the uncertain model \eqref{eq:dynamics:original}. 
Indeed, as solutions of \eqref{eq:dynamics:reformulated} may have unbounded total variation (as in the example presented above), which is not the case for solutions of \eqref{eq:dynamics:original}, 
we make the conjecture that for long horizons $T$, a standard proportional-derivative-integral (PID) controller may better stabilize \eqref{eq:dynamics:original} than \eqref{eq:dynamics:reformulated}, and that similar properties hold for adaptive controllers.

Alternatively, approaches which bound the model error through the Bayesian posterior predictive variance \cite{Khojasteh_L4DC20} or confidence sets holding jointly over time \cite{koller2018,LewEtAl2021} exist. Given these probabilistic bounds, a policy can be synthesized yielding constraints satisfaction guarantees.

\ifCLASSOPTIONcaptionsoff
  \newpage
\fi

\bibliographystyle{IEEEtran}
\bibliography{IEEEabrv,ASL_papers,main}

\end{document}